\documentclass[prl, aps,reprint,superscriptaddress,10pt]{revtex4-1}
\usepackage{graphicx}% Include figure files
\usepackage{dcolumn}% Align table columns on decimal point
\usepackage{bm}% bold math
\usepackage{threeparttable}
\usepackage{gensymb}
\usepackage[english]{babel}
\usepackage{inputenc}  
\usepackage{amsmath} % needed for \tfrac, \bmatrix, etc.
\usepackage{amsfonts} % needed for bold Greek,
\begin{document}
\title{Single photon sources from two dimensional materials and their interfacing with plasmonic waveguides}

\author{Akriti Raj}
\affiliation{ Department of Physics, Birla Institute of Technology, Mesra, Ranchi-835215, Jharkhand, India}
\author{Laxmi Narayan Tripathi*}
\affiliation{ Department of Physics, Birla Institute of Technology, Mesra, Ranchi-835215, Jharkhand, India}
\email{nara.laxmi@gmail.com}

%\dates{Compiled \today}
%\ociscodes{ 220.4241  Nanostructure fabrication;(220.3740) Lithography;  (170.6280)  Spectroscopy, fluorescence and luminescence }
%\doi{\url{http://dx.doi.org/10.1364/ao.XX.XXXXXX}}
\begin{abstract}
We present  a comprehensive review of  formation and  control  of single photon sources from two dimensional materials. We critically analyze the creation and control of these quantum emitters  both at room  and at low temperature.  Deterministic formation and position controlled methodology is elaborated and pictorially explained. We enlist major single photon sources from various two dimensional and non two dimensional resources.  Plasmonic waveguides and the interfacing of single photon sources with them are critically analyzed and reviewed. Furthermore, we envisize the geometry of plasmonic waveguides for maximum coupling of single photons.
\end{abstract}
%\setboolean{displaycopyright}{false}
\maketitle
\section{Introduction} 
 Modern quantum communication devices need non-classical light sources producing photons with well defined and quantum correlation properties. A fundamental resource in this regard is a genuine single photon emitter (SPE) \cite{Aharonovich2016}. The robust and deterministic generation of single-photon states is an essential process required for implementing the vast majority of quantum photonic technologies. One of the most promising Single Photon Source (SPS) is solid state based atom-like fluorescent defects and quantum dots. They promise to combine the outstanding optical properties of atoms and the scalability of solid state systems. Tremendous progress in this field has been obtained in last years by utilizing solid-state single photon sources based on semiconductor quantum dots and atomic defects such as nitrogen vacancies in diamond. However, many fundamental advances in quantum technologies heavily rely on single-photon sources which are stable, bright, and which can be replicated arbitrarily. Recently, single photon emission from defects in inorganic two-dimensional transition metal dichalcogenides (TMDC) single layers such as $WSe_2$ has been demonstrated \cite{Koperski2015}. It furthermore has been shown, that such layered structures can be easily integrated with plasmonic nanostructures via deposition of electrical contacts for pumping the defects in $WSe_2$ \cite{Schwarz2016}. Crucially, it has been outlined that the formation of bright quantum emitters in $WSe_2$, can be induced by the morphology of the substrate via local strain engineering, which holds great promise for their scalable implementation. 
 
 Thus far, monolayer-based single photon sources have been mostly realized by placing monolayers on dielectric substrates \cite{Han2014b}. This architecture, however, intrinsically suffers from low photon extraction efficiencies, since a major fraction (typically $> 90 $) of the photons are leaked into the substrates. Combining monolayer SPSs with resonant cavities, plasmonic antennas or waveguide architectures is still in its infancy yet holds the greatest promise for more profitable exploitation of the approach. Recently, L. N. Tripathi et al. have shown the precise positioning of colloidal quantum dots in the nanogap of silver metamaterials \cite{Tripathi2015}, controlled photoluminescence emission \cite{Tripathi2014} and decay \cite{Tripathi2018a} from monolayers of colloidal quantum dots using gold nanorods (GNR). Further, Kern et al. \cite{Kern2016} have shown precise positioning of strain induced single photon sources between the gold nanorods. The role of plasmonic nanoantenna and electric field enhancement \cite{Gramotnev2014, Zenin2015} of the emission intensity and decay rate of the single photons from TMDC monolayers is an open problem, which will be addressed in this project. Although a theoretical proposal on how to obtain an efficient single photon emission and collection from a hybrid metal-semiconductor assembly via the coupling between a quantum emitter and the excitation of gap plasmons has been recently shown \cite{Lian2015} the corresponding experimental realization is still very challenging and not realized yet. Why plasmonic approaches, rather than dielectric cavities? We know that the mode volume of a dielectric cavity \cite{Koperski2015} $(V) \ge (\frac{\lambda}{2n})^3$ where $\lambda$ is a wavelength of excitation and n is the refractive index of the cavity \cite{Bozhevolnyi2016, Coccioli1998}. Therefore, the diffraction limited ultimate cavity quantum emitter rate is finite, given by $\gamma_{emd}=\sqrt{\omega\gamma_0}$ where $\omega$ is excitation frequency, $\gamma_0$ is intrinsic emission rate of quantum emitter without the cavity. In contrast, Purcell factor for gap surface plasmon resonance antenna is given by $F_p$ $\approx$ $\lim_{t \to 0}$ $(\frac{\lambda_p}{t})^3$ $\to$ $\infty$ which could be infinitely large for infinitely small gap thickness. It is only limited by gap width of the metal-dielectric-metal antenna and the  resonance wavelength. Hence the ultimate emission rate in the presence of plasmonic cavity can theoretically be extremely large \cite{Bigourdan2016}, given by $\gamma_{emp}$ = $F_p$ $\gamma_0$, where, $F_p$ is Purcell enhancement factor in presence of plasmonic antenna. Therefore, a brighter and faster emission can be achieved using a plasmonic nanogap-antenna rather than photonic crystal cavities or other conventional dielectric cavities \cite{Bozhevolnyi2016}.
 
 Furthermore, although the photonic components are superior to electronic components in terms of operating bandwidth, they still suffer from complications associated with miniaturization and high-density integration of the quantum circuits at the nanoscale. Even, diffraction limit in the dielectric waveguides does not allow the localization of electromagnetic waves much smaller than the wavelength of light. Plasmonic waveguides, on the other hand, guide the radiation in the form of strongly confined surface plasmon polariton modes \cite{BOZHEVOLNYI2009, Han2013}. Thus providing a promising solution for manipulating single photon in coplanar structures leading to unprecedented small footprints. Coupling of emission from the single quantum emitter in the channel plasmon has been recently shown with V-groove by the group of Romain Quadrant, S.I.Vojhevolnyi \cite{Bermudez-Urena2015} and Sergey I. Bozhevolnyi \cite{Smith2014}. The efficient coupling of the single emitter to the supported modes of the surface plasmon modes is still an open problem.
 
 \begin{figure*}
 	\centering
 	\includegraphics[width=1.0\linewidth]{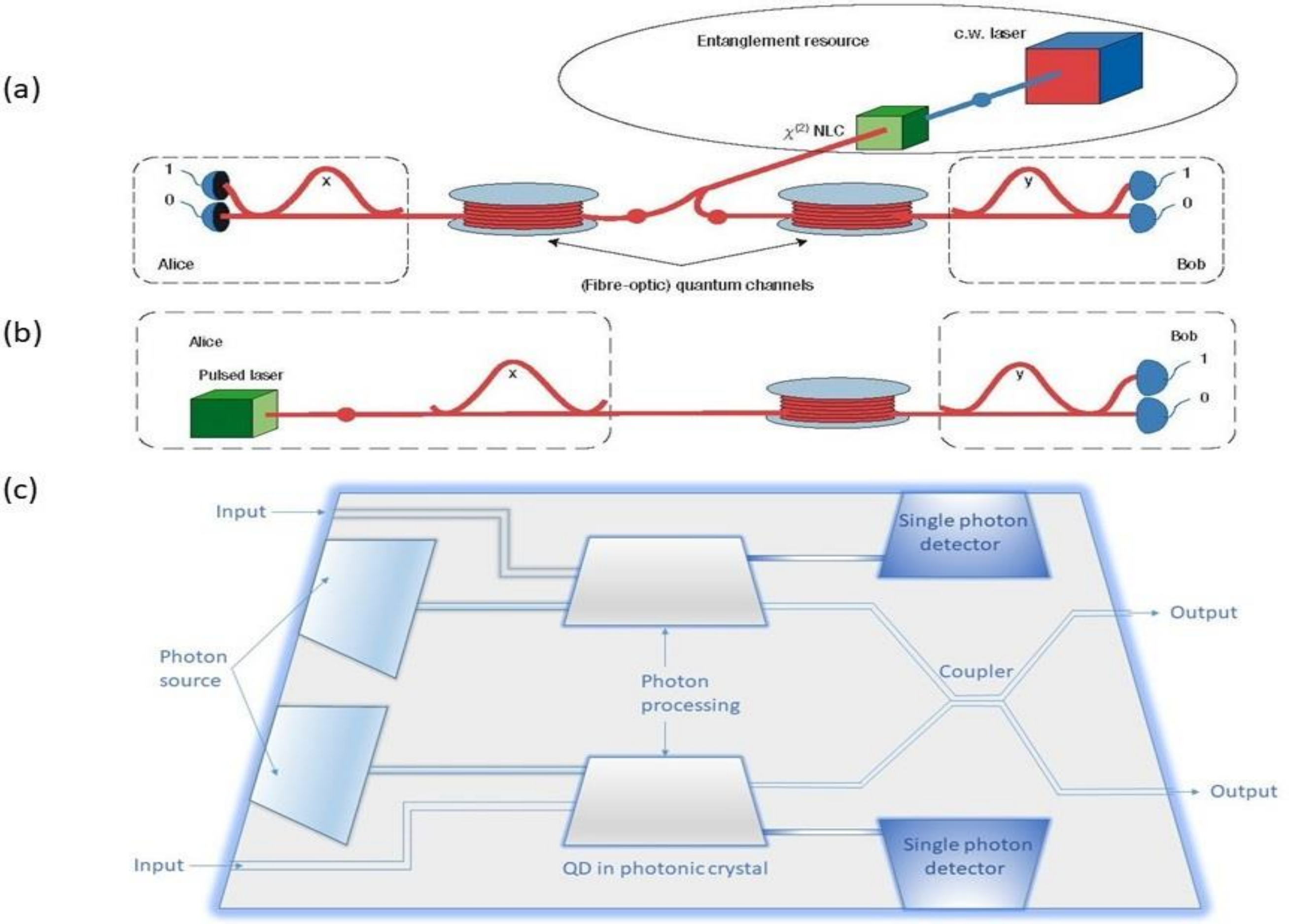}
 	\caption{(a) Schematics of Franson interferometer \cite{Franson1989} used to test the energy-time entanglement of the entanglement resource. CW: continuous wave laser, NLC:  nonlinear crystal and $\chi^{2}$: second order nonlinear crystal susceptibility. Each correlation between each of Alice and Bob$ '$s results 0 or 1 are dependent on both x and y phase measurement settings. Adapted with permission from \cite{Gisin2007}, \textcopyright~(2007) Nature publishing group. (b) In Franson interferometer setup, single heralded photon along with a attenuated pulsed diode laser is replaced in place of entanglement resource. Adapted with permission from \cite{Gisin2007}, \textcopyright~(2007) Nature publishing group. (c) Integrated quantum circuitry and quantum memory. \cite{Aharonovich2016}}
 	\label{fig:fig1}
 \end{figure*}

\begin{figure*}
 	\centering
 	\includegraphics[width=1\linewidth]{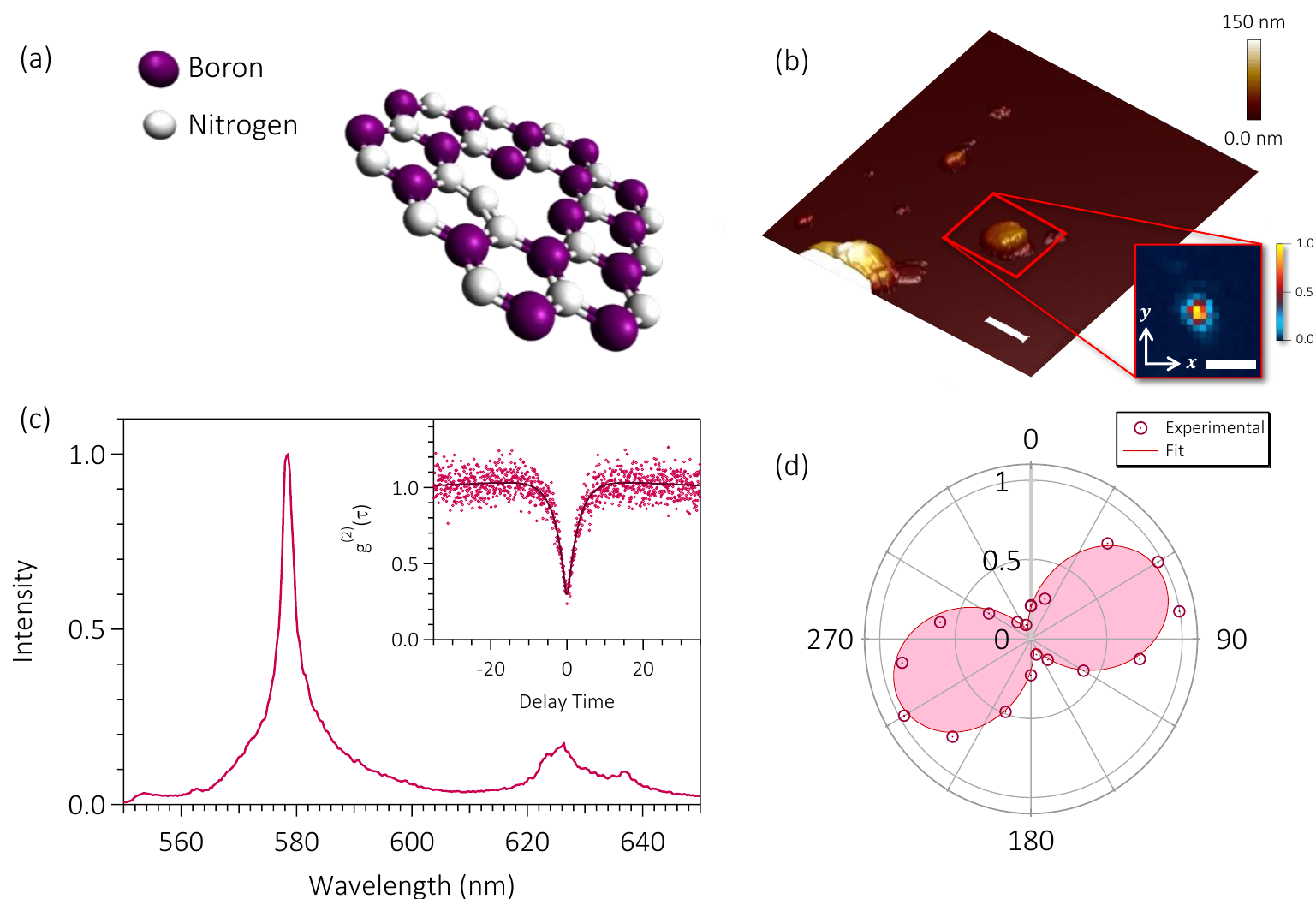}
 	\caption{Optical characteristics of the luminiscent hBN defects. (a) Schematic illustration of
 		the nitrogen-vacancy centre, $N_B V_N$. (b) AFM image showing the morphology of the hBN flake used for plasmonic coupling. At their widest points, the approximate dimensions of flakes are $x=560$ nm and $y=440$ nm. Inset: confocal PL image of the hBN SPE. The scale bar is 1 micrometer (c) Fluorescence spectrum of the hBN SPE. The emitter ZPL is at 578 nm with a full-width at half-maximum (FWHM) of 2.33 $\pm$ 0.04 nm. The inset shows the second-order autocorrelation function is 0.24 indicating single photon emission. (d) Polar graph of emission showing linearly polarized emission. With respect to the north pole of the flake, the linear transition dipole is oriented at an angle of 67$^{\circ} $. The orientation of the north pole of the flake is shown in (a) inset. All optical measurements are performed with 532 nm laser at room temperature. Adapted with permission from \cite{Nguyen2018}, \textcopyright~(2018) Royal Society of Chemistry}
 	\label{fig:fig2}
 \end{figure*}

 With the shift from electronic to quantum devices, these single photon sources prove to have a very promising future in terms of ultra-high processing speed, security, coupling, and quantum efficiency. Several future applications include communication scheme \cite{Lo2014,Scarani2009}, light flux meteorology \cite{Giovannetti2011}, quantum optomechanics and quantum computing \cite{Gimeno-Segovia2015,Kok2007,Aspuru-Guzik2012}. Quantum communication relies on communication at large distances through qubits. Though the theory of quantum communication is a broad field including complexities \cite{Brassard2003} and bit strings \cite{Buhrman2005}, quantum key distribution (QKD) \cite{Bennett1984, Ekert1991, Gisin2002} is one of its most promising applications. The basic thought behind the process is the transfer of information between a sender and a receiver (generically referred to as Alice and Bob respectively) physically separated at a large distance. The information is encoded by Alice and teleported to Bob along with the entanglement-entanglement swapping. In contradiction to Einstein$ '$s theory of local action, Bells theorem successfully explained theoretically as well as practical nonlocality. Nonlocality and entanglement are deeply related and together form the basis of quantum communication. The most obvious degree of freedom to code information is polarization, but polarization can be unstable in fiber cables (birefringence etc). Another way is to use the energy-time entanglement \cite{Franson1989} analogs to EPR theory. Figure \ref{fig:fig2}(a) shows the photon from continuous wave laser is being down-converted to a pair of photons by passing through a nonlinear crystal \cite{Tanzilli2001} each sent to Alice and Bob. These photons possess uncertainty in time and energy, as they hit the detector at the same time they tend to produce an interference pattern. This configuration is suitable for quantum key distribution (QKD) but a much-simplified version (Fig.~\ref{fig:fig2}(b)) as shown in the other figure  is used in practice. The continuous wave (CW) laser and nonlinear crystal can be replaced by a standard telecom laser diode without compromising the security \cite{Inamori2007, Gottesman2004} of the system. This operation has a mean photon number per pulse of $<n> \approx 0.5$. For an ideal single-photon source $ <n> = 1$. QKD using a true SPS second order auto-correlation function at zero delay, also known as purity of single photon, ($ g^2(0)$) = $10^{-3}$ for fibre length of more than 120 km has already been demonstrated \cite{Schmitt-Manderbach2007,Lo2005,Takemoto2015}. The raw and secure key rates at 100 km were measured to be $\approx $ 80 and $\approx $ 28 bits per second. This order is still 1-2 magnitude smaller than what is realized using a standard attenuated laser diode, provoking the need for improvement of SPS (improved brightness and extraction efficiency). Distant quantum nodes often have repeaters in between to compensate for the loss. It has been shown that millions of high-efficiency SPS will be needed per repeater in order to produce a required statement that can beat the limits of repeater fewer QKD \cite{Pant2017}.

 In this report, we survey  the state of art and types of single photon sources both based on transition metal dichalcogenides(TMDC) and non-TMDC.  We review the deterministic generation and position of these sources both at room temperature and at low temperature (4.5 K). Elaborate discussion  of strain induced  single photons  from  the two dimensional materials is followed by a comprehensive summary table of  major types of single photon sources. We then introduce plasmonic waveguides and coupling of single photon sources with various plasmonic waveguides. We also shed inputs on geometry of the wave guides for maximum coupling of single photons for future applications of quantum information processing.

 \section{TMDC as single photon source}
 Two-dimensional materials have been one of the most studied material. The successful exfoliation of graphene and their display of unique properties as compared to the bulk have proved to be very promising and inspired several others to further study the two-dimensional materials in details. As SPE$ '$s, these monolayer  are Van der waall crystals when compared to the three-dimensional material have an advantage of less total internal reflection and easy coupling with interconnects that increases their integrability and scalability \cite{Butler2013, He2015}. It has been shown that monolayer transition metal dichalcogenides (TMDC) such as $MoS_2$, $MoSe_2$, $WS_2$, $WSe_2$ and hexagonal boron nitride (hBN) are good candidates for single photon source. In TMDC the SPE is associated with localized excitons due to defects whereas in hBN it is due to nitrogen-vacancy defects deep within the band gap. These defects in hBN are one of the brightest SPS in visible spectra detected so far \cite{Tran2016a}. Along with room temperature stability hBN possess some excellent properties such as short excited state lifetime, narrow linewidth, absolute photon stability, and high quantum efficiency. TMDC monolayer turns out to have a direct optical band gap in visible and near IR region that is an added advantage. The emission appears on the edge of the atomically thin flakes due to the creation of a local potential well that traps the exciton and have excellent stability at low temperature \cite{Tonndorf2015}. If we somehow control the edge imperfections we can increase the applications of these SPE$ '$s \cite{Koperski2015}. Research has demonstrated that SPE $WSe_2$ shows a narrow line width, strong antibunching property, and two non-degenerate transitions that are cross linearly polarized. The emission properties can also be controlled by application of external DC electric and magnetic field \cite{Chakraborty2015}. 
 
 \section{Plasmonic waveguides}
 Coupling  a single photon source to a waveguide faces challenge of transport of a single photon over a large distance with minimum loss. An early approach to this problem was the use of metals separated on a nanometer scale or V-shaped grooves on metal. Both of which shared a common problem of high loss and low propagation distance. Then several other options like photonic crystal \cite{Thylen2004}, Silicon on Insulator (SOI) nanowires with ultra-high index contrast \cite{Tsuchizawa2005,Almeida2004}, plasmonic waveguides \cite{Goto2004, Charbonneau2005, Zia2004, Wang2004, Tanaka2003, Kusunoki2005, Pile2004, Bozhevolnyi2006, Pile2005, Xiao2006, Liu2005, Veronis2005}, L shaped gap surface plasmon waveguide \cite{Gramotnev2012} were explored. Out of which the most extensively used method is plasmonic waveguides formed as a result of small separation between metal and high refractive index material. Due to this small separation surface plasmons are generated that gets coupled with the photon resulting in a capacitor like effect and facilitate transportation \cite{Oulton2008}. It had been shown that the excitons can couple effectively to the modes at the center of the plasmonic waveguide, their coupling efficiency and Purcell factor show considerable increases with precise positioning of the emitters. Plasmonic waveguides created by the gap between the monocrystalline silver nanowire and monocrystalline silver flakes using the nitrogen-vacancy (NV) center in nanodiamonds (stable at room temperature) was studied by Sailesh Et al. \cite{Kumar2018}. They demonstrated that the decay rate enhanced up to $\approx$ 50, and efficient channeling of photon up to 82 $ \%$ into the waveguide. In another work of his, the group has demonstrated strong polarization and increase single photon purity when a nitrogen vacancy diamond is placed between two monocrystalline silver cube in dimer configuration \cite{Andersen2017} at a nanometer scale. Coupling of these NV center diamonds has also been done with silver nanocubes \cite{Andersen2016}. Plasmonic have a clear advantage over the conventional components which are wavelength limited, as they offer deep subwavelength confinements and the integration limit expands to the nanometer scale. Research has already demonstrated that nanowires and dielectric waveguide can couple both free as well as localize excitons, the binding energy of which is subject to the dielectric environment (nm scale) and the presence of strain. Properties of mechanically exfoliated $WSe_2$ transferred onto plasmonic waveguide is well explored \cite{Bahk2018}. Theoretical studies predict that if we have a tip at one end of the waveguide then propagating plasmons are slowed down and never actually reach the end. This property will have wide range applications in nano-optics.
 
 \begin{figure*}
 	\centering
 	\includegraphics[scale=0.50]{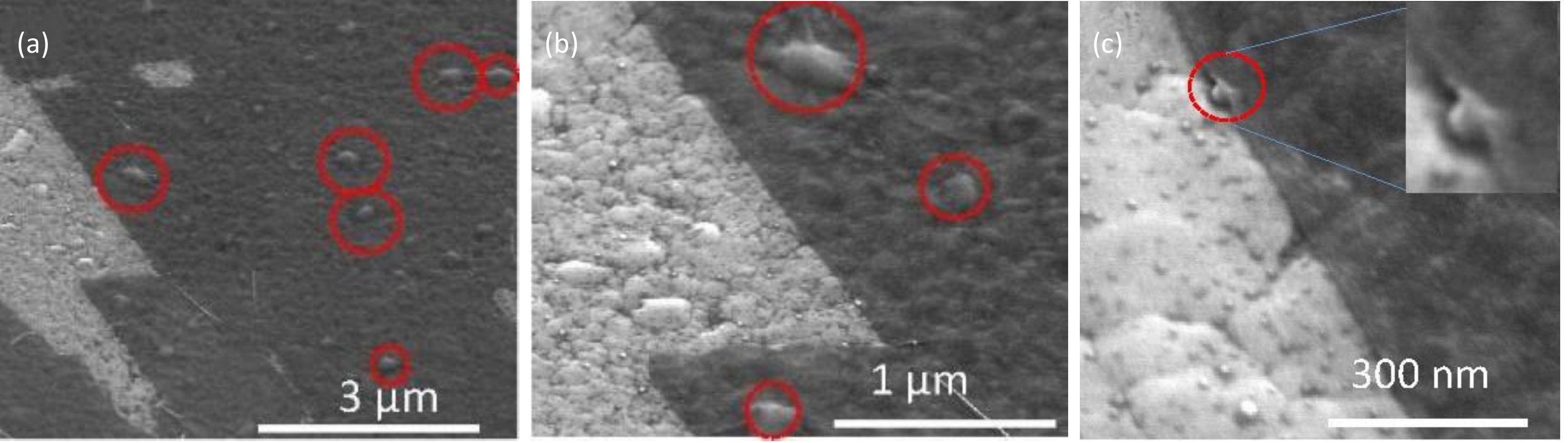}
 	\caption{(a-b) High resolution SEM images showing conformal morphology of
 		$WSe_2$ and silver surface. Red dotted circles show strained monolayer around such nanoparticles. (c) A monolayer is seen to bend over a silver nanoparticle (dotted red circle and inset:  zoomed in picture of the strained monolayer due to the nanoparticle).  Adapted with permission from \cite{Tripathi2018}. \textcopyright~(2018) American Chemical Society.}
 	\label{fig:fig3}
 \end{figure*}
 \begin{figure}
 	\centering
 	\includegraphics[width=1\linewidth]{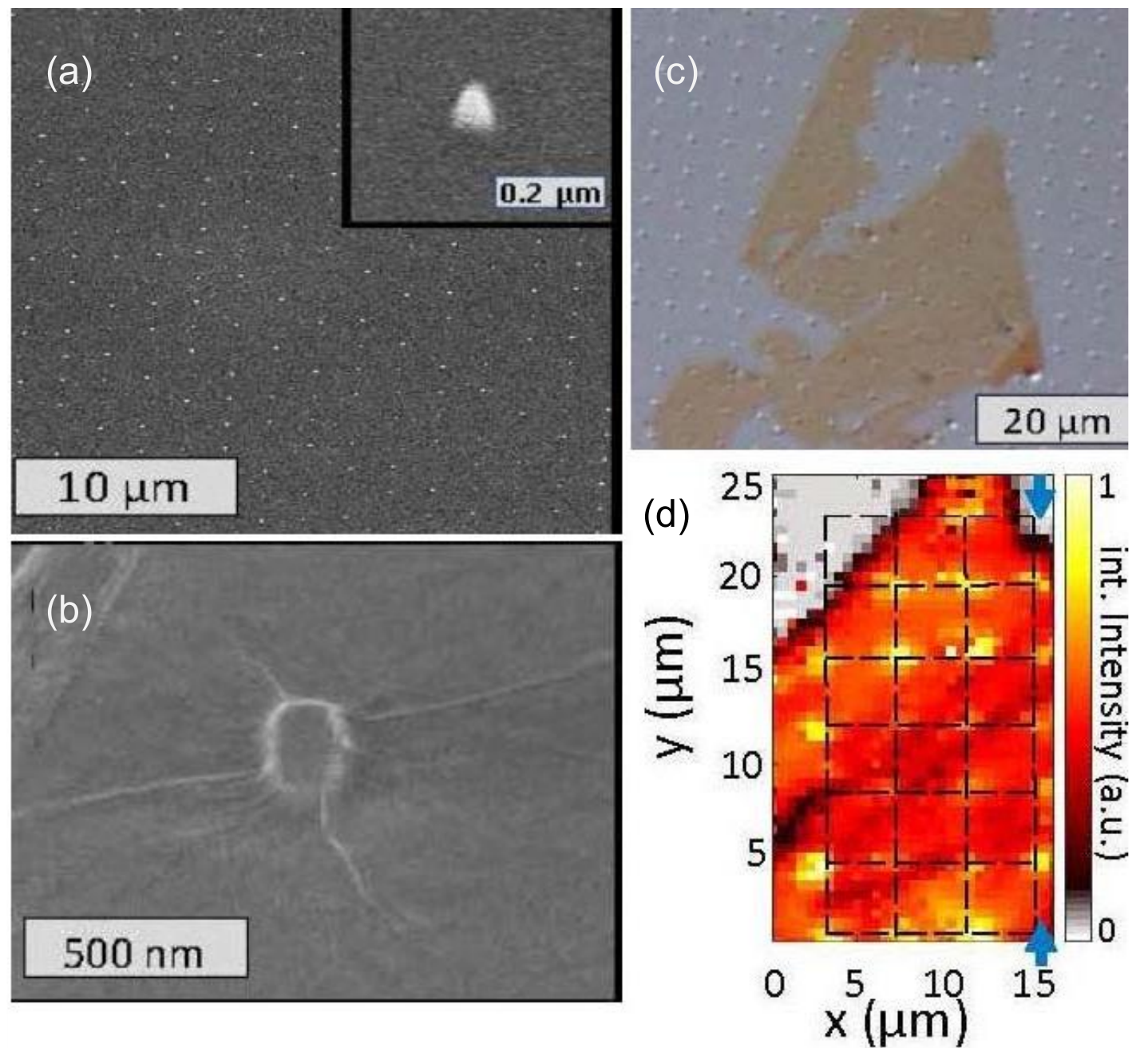}
 	\caption{(a) SEM image of the sample consisting of quantum emitter seeds (metallic nanopillars) and plasmonic nanocavities. Inset: close-up
 		view of a nanopillar. (b) Optical image of the pillar array after successful dry-transfer of an atomically thin $WSe_2$ monolayer. (c) Close-up SEM image of a single pillar, showing the wrinkle formed on the strained monolayer,  covering the pillar. (d) Spatial map of the nanopillar array covered by a $WSe_2$ flake. It also shows the integrated intensity ranging from 700 - 800 nm. The black pattern shows the coinciding of enhanced PL with 4 $\mu$m pillar distance.
 		(black pattern). Adapted with permission from \cite{Iff2018}, \textcopyright~(2018) Optical Society of America. }
 	\label{fig:fig4}
 \end{figure}
 \begin{figure}
 	\centering
 	\includegraphics[scale=0.75]{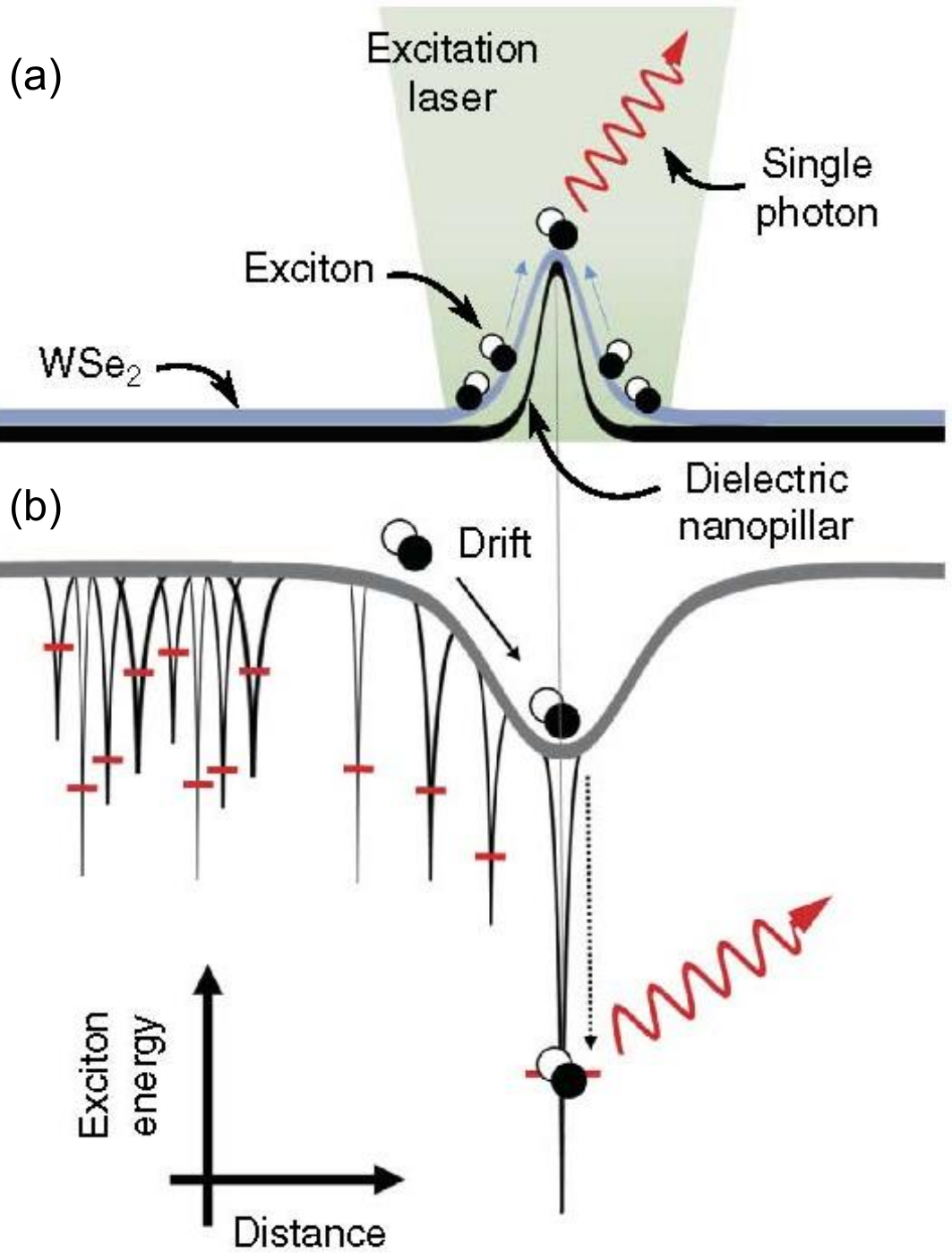}
 	\caption{(a) A point-like elastic strain perturbation is achieved due to the deformation of atomically thin $WSe_2$ by nanopillar. (b) The $WSe_2$ bandgap is modulated by local strain. Randomly distributed localized excitons are superimposed on artificially modulated exciton energy. Efficient funneling of optically created excitons to an individual strain localized exciton trap at the nanopillar center, resulting in a single highly efficient quantum emitter. Adapted with permission from \cite{Branny2017}, \textcopyright~(2017) Nature Publishing Group.}
 	\label{fig:fig5}
 \end{figure}

 \section{Strain induced single photon source}
 Aharonovich et al. \cite{Nguyen2018} demonstrated the coupling of gold nanosphere with a two-dimensional hexagonal boron nitride layer (Fig.~2(a)). The higher the overlap between the source emission and cavity resonance, the better enhancement rates are achieved. And better overlapping can be achieved if the spatial resonance spectra are broad as in the case of nanoantenna \cite{Koenderink2017}. The problem of precise alignment between the emitter and the nanosphere is solved by using an Atomic Force Microscope (AFM) which provides positioning with nanometer accuracy. The pre-characterized hBN is used and nanospheres are manipulated using AFM (Fig.~2(b)), bringing them gradually close to the emitting layer. Initially, a single sphere is positioned for optimum coupling resulting in a single particle confinement configuration. Later a second sphere is positioned in a similar manner forming a double particle configuration. In double gold nanosphere arrangement hot spot is created between the two spheres, a decrease in enhancement is observed with increase in gap size. The second order correlation function is measured for all the three configurations i.e. pristine state, single confinement and double confinement to ensure that the quantum nature is not disturbed. Results show that the coupling significantly increases the emission rates and ultrahigh brightness. The smaller the size of the hBN nanoflakes better positioning can be achieved inside the hot spot. An optical bandpass filter was used to isolate the Zero Phonon Line (ZPL)  (Fig.~2(c)) from any background emissions and the measured value of second-order correlation function was well within 0.5, the degree of polarization (Fig.~2(d)) also increased which can be advantageous for quantum technological applications. As the plasmonic fields are more intense near the metal surface, local excitation enhancement is dominant here and decrease with the increase in distance from metal. The quantum emitters which are perpendicular to the metal surface show more enhancement that the ones that are parallel to it. In double gold nanosphere arrangement hot spot is created between the two spheres, the enhancement decreases with an increase in gap size. The emission lifetime decreased due to the increase in local state density as the electric fields compactly held. This hybrid plasmonic system demonstrated a high count rate with a moderate Purcell enhancement and is reported as one of the brightest emission at room temperature.

 Quantum emitters show an increase in brightness, optical nonlinearities, and emission rate when they are effectively coupled to plasmonic nanostructures \cite{Henzie2009, Hoang2015, Hoang2015a, Waks2010}. But the major problem is the precise alignment between the emitter and the plasmonic nanostructure. One way to achieve this to use the tip of the AFM as described in the previous mention work but the setup is very complicated and is not scalable \cite{Nguyen2018}. Another random way is to place the emitter on an array of plasmonic nanostructure \cite{Tran2017} so that at least one mode is coupled to the emitter. Tao Cai et al \cite{Cai2018} coupled $WSe_2$ to lithographically designed nanopillars. These nanopillars had a silicon base, gold layer and a  $Al_2 O_3$  layer on the top that acts as a buffer \cite{Anger2006}. They numerically calculated the scattering spectrum and showed that Purcell factor enhancement shows a strong dependence on position and wavelength of the emitter. Second order correlation function studies has been done to prove that the emission occurs from a quantum light source. A vivid explanation of the fabrication technique has also been presented.
 
 Among all the SPS TMDC has been in the spotlight, due to their unique property of band tunability by application of localized strain (Fig.~4). Although the exact cause of emissions in TMDC is still unknown extensive studies have been done to manipulate and control the emissions \cite{Tonndorf2015, Koperski2015, Srivastava2015, He2015, Chakraborty2015}. By studying the Raman and photoluminescence spectra of few-layered $WSe_2$ an indirect to direct transition has been recorded \cite{Desai2014}. The applied force from the tip of an AFM can release the pre-strained material \cite{Park2016}. It has been predicted that with the strain the shift in bandgap is of the range of $-90 \frac{meV}{\%} $ for a monolayer $WSe_2$ \cite{Johari2012, Amin2014}. Schmidt et al. \cite{Schmidt2016} presented the absorption spectra of $WSe_2$ under reversible uniaxial strain. The monolayer of $WSe_2$ was transferred onto a polycarbonate substrate and an elastomer coating was applies on the top. The elastomer ensured that the $WSe_2$ flakes remain intact when the strain is applied. The substrate was then placed between the translation stages, the distance between them was gradually reduced. A reversible uniaxial strain was thus experienced by the $WSe_2$ monolayer given by $\epsilon =\frac{h}{2R}$ where h is the thickness and R is the radius of curvature of the substrate. By studying the transmission of light the absorption spectra are measured. Excitons show a shift in resonance energies with an increase in uniaxial strain whereas their binding energy remains independent of the applied strain. Thus, the optical property of the monolayer can be effectively tuned. No appreciable hysteresis is observed on reversing the strain. Spatial dependent studies show that the strain is not uniformly transferred from substrate to the monolayer and the adhesion area between the two play a very crucial role.  Shear stress is developed due to Van der Waal adhesion force between the monolayer and the substrate and strain energy is stored in it. At the monolayer center exciton energy is shifted whereas at the edges the strain energy is released. A lifting force is thus experienced by the edges, resulting in a very small shift of exciton energy. The strain gradient developed within the monolayer depends on the ratio of Van der Waal force (i.e. the adhesion force) and Young$ '$s modulus of the monolayer.

 	\begin{table*}
 		\begin{threeparttable}
 			\caption { \label{tab1} \emph{
 					\textbf{Summary of quantum  properties of single photons sources based on both Transition metal dichalcogenides (TMDC) and Non Transition metal dichalcogenides (Non TMDC).}} }
 			\tabcolsep=0.08 cm
 			
 			\begin{tabular}{lllllllllcccccccccccccccccccccccccccccccccccccc}
 				& \cr
 				\hline
 				SPS &  &  &   & Quantum & materials &  & &  &  &\cr
 				
 				Properties \tnote{2,3} &  &  &  & NON & TMDC &   & & & TMDC& \cr                
 				
 				\hline
 				& CNT \tnote{1} &N-QD \tnote{1}  &SiC \tnote{1} & Ar-QD \tnote{1} & SiVD \tnote{1}  &ZnO \tnote{1} & BN \tnote{1}& NVD & &\cr
 				
 				&  &  &   &  &  &  & &  &  &\cr
 				
 				Lifetime & 0.131$\pm$ 0.043 & 0.357 at 220K  &0.81$\pm$0.01 & 0.520 & 1  &1-4 & 2.4& 12-22,18 &1-3,2.4 &\cr
 				(ns) & \cite{Ishii2018} & \cite{Wang2017}  &\cite{Wang2018} & \cite{Sapienza2015} & \cite{Aharonovich2016}  &\cite{Morfa2012} & \cite{Grosso2017}& \cite{Aharonovich2016, Schroeder2011} &\cite{He2015, Cai2017} &\cr
 				
 				&  &  &   &  &  &  & &  &  &\cr
 				
 				Count rate & $10^{5}$ &   &$10^{6}$ & $10^{7}$ & 4.8 $\times$ $10^{6}$  &$10^{6}$ & 7 $\times$ $10^{6}$& 2.4$\times$$10^{6}$ &3.7$\times$$10^{5}$\cite{He2015} &\cr
 				range (Hz) & \cite{Ishii2018} &   &\cite{Wang2018} & \cite{Sapienza2015} & \cite{Neu2011}  &\cite{Morfa2012} & \cite{Grosso2017}& \cite{Schroeder2011} &,3 $\times$$10^{4}$\cite{Cai2017} &\cr
 				
 				&  &  &   &  &  &  & &  &  &\cr
 				
 				Purity & 0.098, 0.2 \tnote{4} & 0.13 at 4.7K  &0.05$\pm$0.03 & 0.009$\pm$0.005 &   & & 0.077& 0.16-0.3 &0.25 \cite{Tripathi2018}  &\cr
 				$g^{(2)}(0)$ & \cite{Ishii2018} & 0.21 at 220K  &\cite{Wang2018} & \cite{Sapienza2015} &   & & \cite{Grosso2017}& \cite{Schroeder2011} &0.17$\pm$0.15 \cite{Iff2018} &\cr
 				&  & \cite{Wang2017}  & &  &   & & &  & &\cr
 				
 				&  &  &   &  &  &  & &  &  &\cr
 				
 				Temperature & RT\cite{Ishii2018} & RT \cite{Wang2017} &RT\cite{Wang2018} & 4K \cite{Sapienza2015} & RT \cite{Aharonovich2016} &RT \cite{Morfa2012}& RT\cite{Grosso2017}& RT \cite{Schroeder2011} & 4K\cite{He2015} &\cr

 				&  &  &   &  &  &  & &  &  &\cr
 				
 				Integrable & Si  & Dielectric  & & Dielectric & Dielectric  &Dielectric & Photonic& Dielectric & Dielectric &\cr
 				& microcavity & Plasmonic  & & Plasmonic & Plasmonic  &Plasmonic & circuit& Plasmonic &Plasmonic &\cr
 				& \cite{Ishii2018} & \cite{Michler2009}  & & \cite{Sapienza2015} & \cite{Neu2011}  &\cite{Morfa2012} & \cite{Grosso2017}& \cite{Schroeder2011} &\cite{He2015, Cai2017, Wang2018a} &\cr
 				&  &  &   &  &  &  & &  &  &\cr
 				\hline
 			\end{tabular}
 			
 			\begin{tablenotes}
 				\item[1] Abbreviations- carbon nanotube (CNT), Nitride Quantum dot (QD), Arsenide Quantum dot (QD), Silicon Carbide (SiC), Silicon vacancy diamond(SiV), nitrogen vacancy diamond (NV), Zinc Oxide (ZnO) and Boron Nitride (BN).
 				\item[2] All the important properties such as the lifetime of the emitter in a nanosecond, count rate (number of the single photon produced per second), Purity of the single photon source measured in terms of the second order correlation function at zero time delay ($ g^{2}(0) $). The working temperature (RT$\equiv$ Room Temperature) been practically associated with a plasmonic or dielectric to further enhance its properties are reported in the table.
 				\item[3] Note that all the values quoted here are approximate, and may slightly vary from one work to another. A detailed explanation for the given values can be seen in the references cited.
 				\item[4] 0.098 is with filter and 0.2 is without a filter.
 				
 			\end{tablenotes}
 		\end{threeparttable}
 	\end{table*}

 \begin{figure}
 	\centering
 	\includegraphics[scale= 1.0]{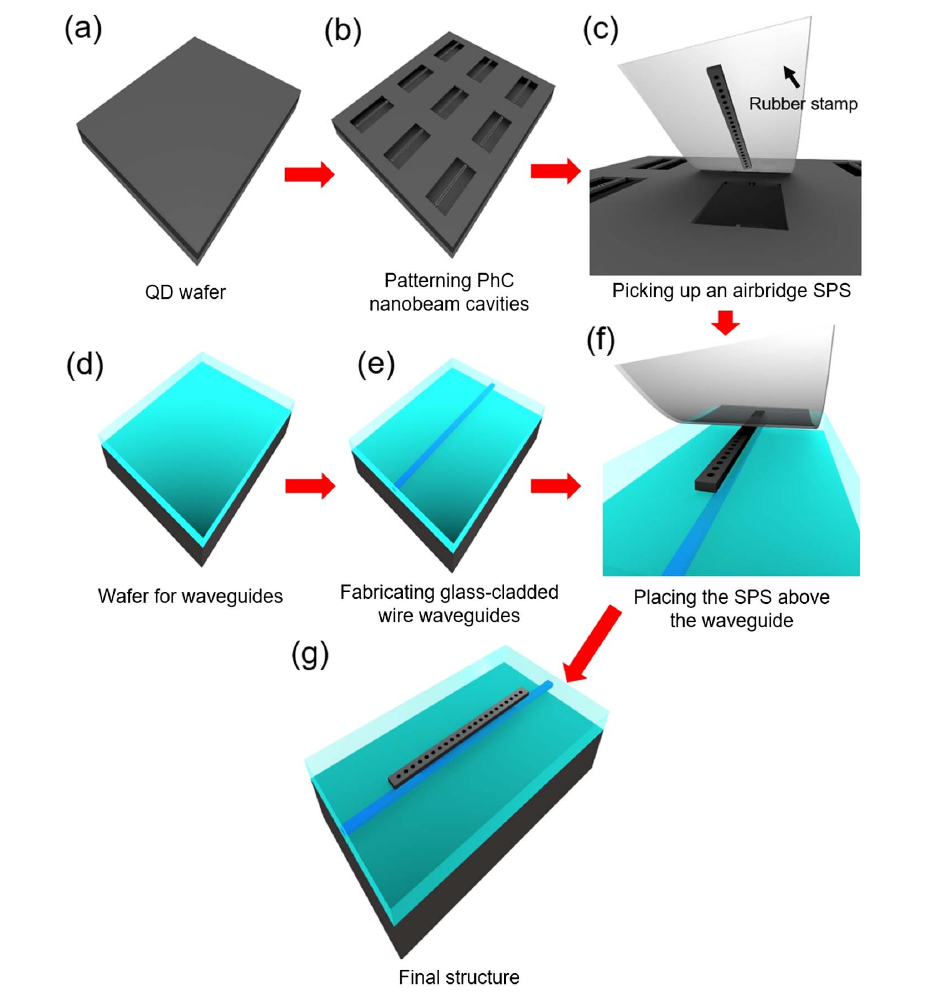}
 	\caption{Schematics of transfer printing. (a)Prepare SPS on a QD wafer,
 		by patterning (b) an array of PhC nanobeam cavities. (c) An airbridge
 		SPS is picked up by attaching and quickly peeling off the transparent rubber stamp. Meanwhile, we use (d) another wafer for preparing
 		(e) glass cladded wire waveguides. (f)The SPS is placed above the waveguide and then slowly releasing the stamp.
 		(g) Final structure. Adapted with permission from \cite{Katsumi2018}, \textcopyright~(2018) Optical Society of America.}
 	\label{fig:fig6}
 \end{figure}
 
 \begin{figure*}
 	\centering
 	\includegraphics[scale=0.60]{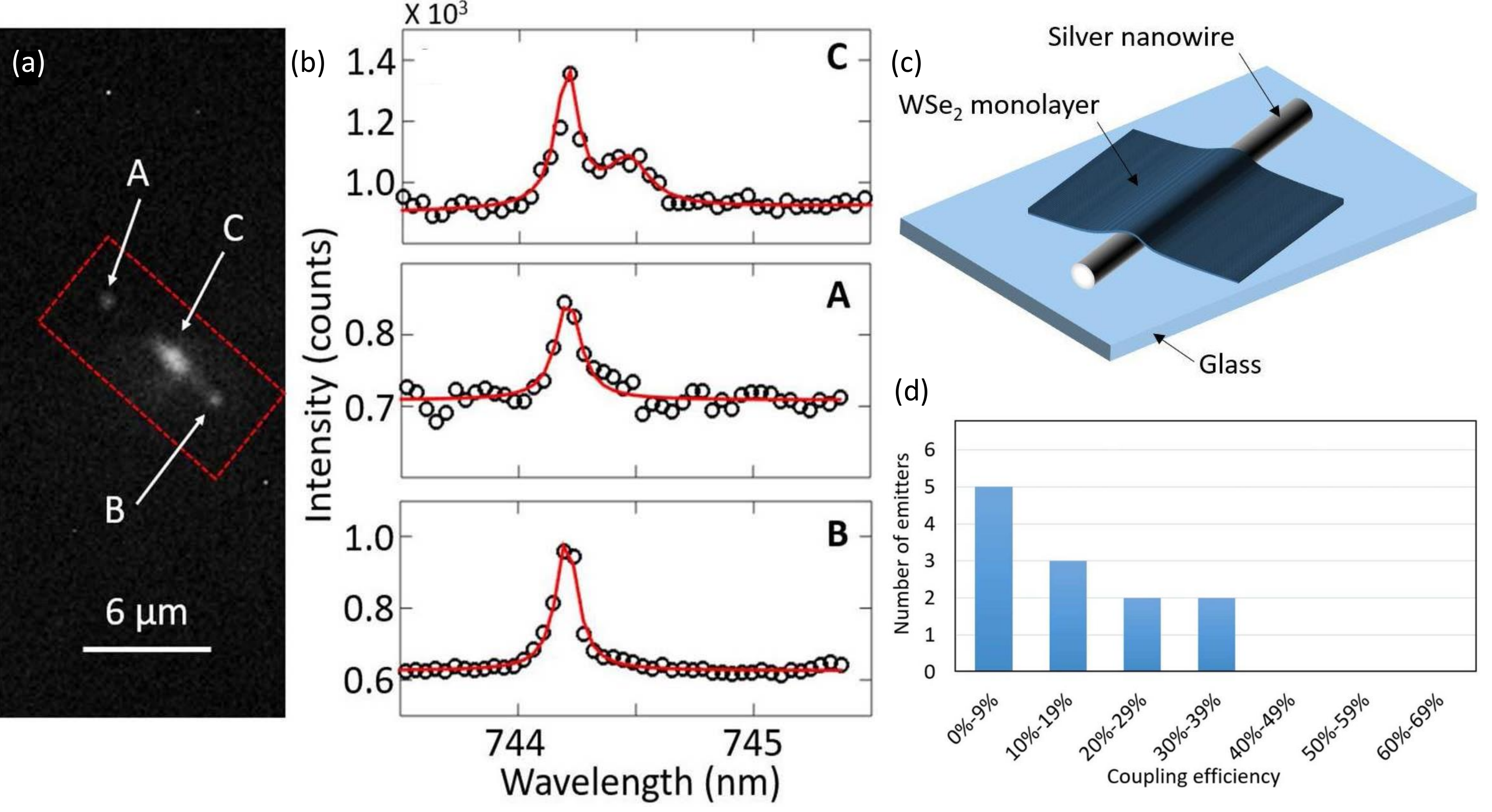}
 	\caption{(a) A photoluminescence intensity map showing the emission at the emitter (C) and at both ends A and B of the silver nanowire. The position of the silver nanowire is indicated by red box. (b) Photoluminescence spectra collected at the emitter "C" (top) and on both sides "A" (middle) and "B" (bottom) of the siver nanowire. (c) 3D schematic layout of a silver nanowire/$WSe_2$ monolayer device. (d) Distribution of the coupling efficiencies of single-defect emitters to the silver
 		nanowires. Adapted with permission from \cite{Cai2017}, \textcopyright~(2017) American Chemical Society.}
 	\label{fig:fig7}
 \end{figure*}

 \begin{figure}
 	\centering
 	\includegraphics[scale = 0.75]{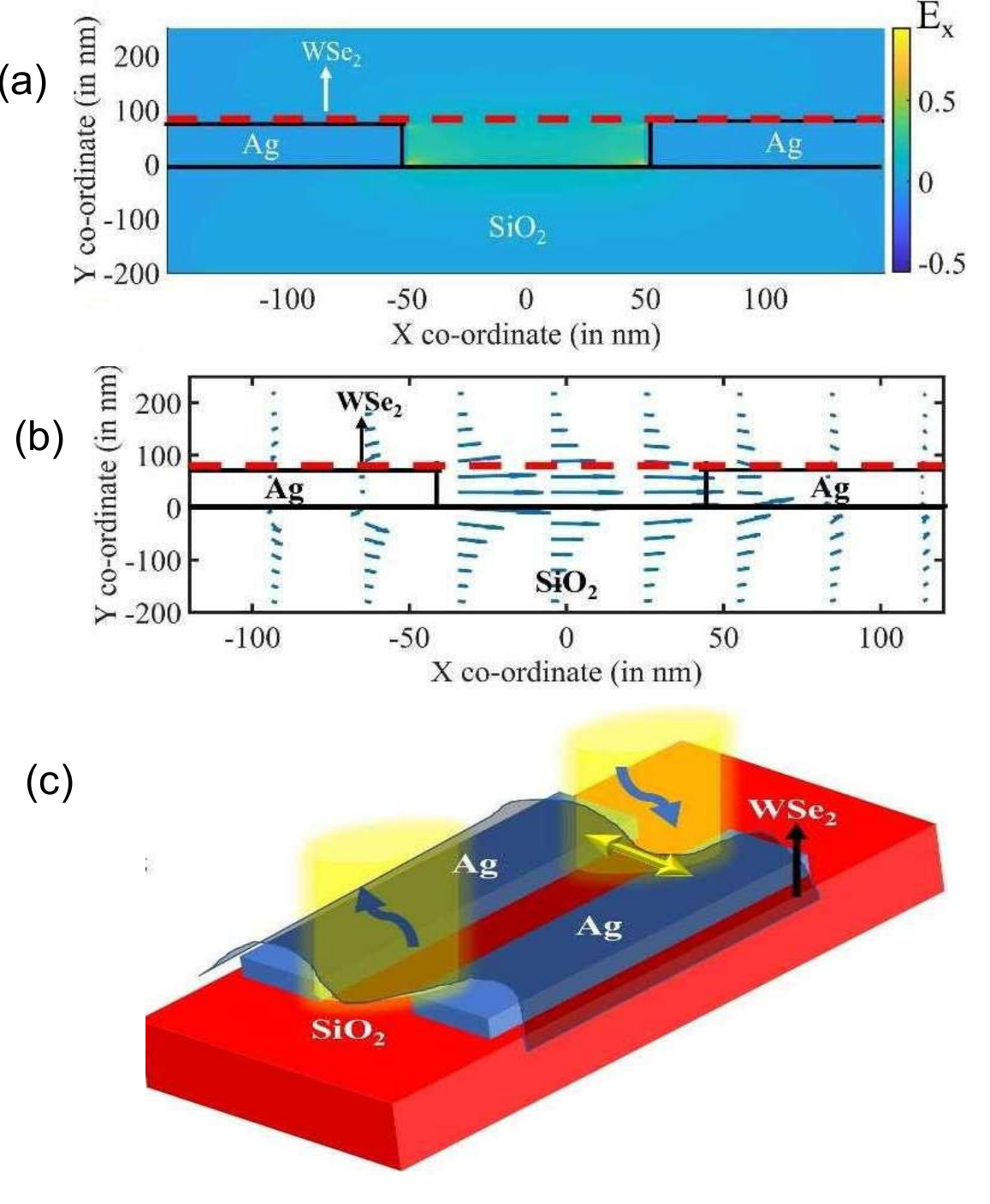}
 	\caption{Finite difference eigenmode simulation (Lumerical Inc.) of a MIM waveguide (a) showing the x-component of the propagating (along z-axis) electric field for plasmon mode, and (b) showing the vector-plot of the electric field in the gap region. (c) Schematic of a MIM waveguide covered by $WSe_2$ monolayer. The quantum emitter in $WSe_2$ is represented by a yellow dipole and the blue arrows indicate the excitation and collection points. Reproduced from \cite{Dutta2018}, with the permission of AIP Publishing.}
 	\label{fig:fig8}
 \end{figure} 
 
 L N Tripathi et al. \cite{Tripathi2018} studied monolayer $WSe_2$ on a rough metallic surface (silver), the roughness helped in strain-induced single photon emission (Fig.~3(c)) and the metal surface ensured the increase in the spontaneous rate of emission. A strain induced enhancement in emission intensity is observed using SEM (Fig.~3(a-b)). Further detailed study of these quantum dots like emitters was done by studying their linewidth dependence on counts and excitation power. It is found that the line width is almost constant with increase in excitation power, so the system can work effectively for broad power. For a single quantum dot wavelength Vs time studies show some spectral wandering that contributes to the linewidth broadening of the emitted radiation. The emission does not blink or bleach and show strong linear polarization. Second order correlation function studies demonstrate an anti-bunching feature at zero time delay and some bunching at finite time delay. Time-resolved decay dynamics studies show an enhanced decay time up to nanoseconds attributed to the formation of nanoantennas \cite{Novotny2011, Biagioni2012} leading to increases in photonic mode density. When the orientation of field and dipole is same this enhancement in radiative decay filed is observed. Further analysis is done by manipulating the size of the cone. For a fixed cone angle and increasing, the height absorption spectrum is blue shifted, for fixed height and decreasing base length absorption spectrum is red shifted. For hemispherical silver nanoparticle with decreasing base diameter, the spectrum is again redshifted. And it is verified that the coupling is maximum at the base of the cone rather than at the tip.

 Branny et al. \cite{Branny2016} used $MoSe_2$ monolayer flakes extracted through viscoelastic stamping \cite{Castellanos-Gomez2014}. Three samples are made out of which two are exfoliated on a gold substrate creating some wrinkles for induced strain control demonstrations and the third sample is kept between two electrodes Cr/Au and n dopes Si substrate to study magneto-optic effects onto the sample. A magnetic field is applied perpendicular to the sample. Photoluminescence spectral studies are done using a Si CCD. Sample 1 and 2 show charged and neutral excitations peaks resulting from good quality flakes. Detailed emission spectra study show that the peaks are not shifted apart from some redshift at the strained areas. Unlike $MoS_2$, the emission peaks are not substrate dependent. The emission lines are very closely spaced making them difficult to extract. Some linewidth broadening is observed due to fluctuations in the environment because of the presence of gold substrate. It can be concluded that, similar to $WSe_2$ \cite{Tonndorf2015,Kumar2015} emissions in $MoSe_2$ are also sharp and strain localized. In the second section photoluminescence studies are done as a function of applied magnetic fields. With an increase in field amplitude Zeeman splitting increases and from linear fitting, the data g factor can be calculated. The calculated g factor is close to trion and neutral exciton for monolayer $MoSe_2$ \cite{Li2014, MacNeill2015, Wang2015} implying some relation between a discrete emitter and the once formed by a carrier at the discrete gap at k point. Contrast to this the g factor valve obtained for $WSe_2$ is very large as compared to the local exciton \cite{Koperski2015,Srivastava2015,He2015,Chakraborty2015}, inspiring further studies \cite{Wang2015,Srivastava2015,Aivazian2015,Stier2016} in this direction. Another quantum dot show zero filed splitting and proportional to field amplitude as $\sqrt{\delta_1 ^2 + \triangle_Z ^2} $, where $\delta_1$ is the fine structure splitting and $\triangle_Z$ is the Zeeman splitting. The large value of $ \delta_1$ implies strong Coulomb interactions. Further analysis is done by tuning the ratio between trion to neutral exciton by applying an external bias. This will help in producing modified emissions that proves to be useful for optical and electrical manipulation of single spin and quantum optics application.

 White et al. \cite{White2018} integrated $WSe_2$ monolayer with a lithium niobate directional coupler. TMDC is very convenient to fabricate as the monolayer can be placed on any substrate without worrying much about any complex bonding process \cite{Palacios-Berraquero2016, Brotons-Gisbert2018, Tonndorf2017, Frisenda2018}. Lithium niobate is useful for photonic telecommunication due to a broad wavelength spectrum, low transmission loss, and fast switching capabilities. Their properties have already been exploited to build waveguides, beam splitters, ring resonator \cite{Krasnokutska2018} both in Ti-diffused and ridge structure \cite{Hu2007}. Using photolithography waveguide was printed on a $LiNbO_3$ wafer and Ti was diffused onto it, in a wet oxygen environment. Strain engineered \cite{Kumar2015} monolayer $WSe_2$ flakes were transferred onto the waveguide. A HeNe laser was sent from the opposite end of the waveguide insuring deterministic transfer of flake on the optical modes. The photoluminescence spectra were collected from the output end of the directional coupler. Some strong emission lines were observed corresponding to quantum dot emission from the $WSe_2$ flakes. On-chip Hanbury Brown Twiss set up was demonstrated and it was confirmed that directional coupler can act as the beam splitter.

 Though development in the field of single-photon sources and plasmonic circuits and excelled independently, their integration is still limited \cite{Davanco2017, Elshaari2017}. Katsumi et al. \cite{Katsumi2018} presented a simple scalable approach of transfer printing \cite{Menard2004, DeGroote2016, Justice2012, Yang2012, Lee2017} to couple SPS with the existing plasmonic circuits prepared independently and with a very high coupling efficiency (Fig.~6). The previous approach to this problem includes joint fabrication of the SPE and the waveguide or using a micromanipulator \cite{Kim2017} both of which have very less coupling efficiency. Two QD wafers are taken one for the fabrication of nanobeam cavity (air bridge SPS) and another for the fabrication of waveguide. This wire waveguide is buried inside the glass cladding. They then used a rubber stamp to transfer one SPS from the first processed wafer to the processed waveguide under an optical microscope. By carefully removing the stamp SPS can be placed on the waveguide. This simple pick and place transfer technique demonstrate very high coupling efficiency. Q factor, cavity waveguide coupling, emitter cavity coupling all show dependence on cavity waveguide distance. Even in the case of some misalignment, the high coupling efficiency is still maintained. They have demonstrated all the measurements on glass-clad GaAs structure and spin on glass coating. The photoluminescence spectra for this pair was measured and all the results were verified. Later on, this method was also checked to couple multiple SPE into the waveguide for validating the scalability of the process.
 
 Oliver et al. \cite{Iff2018} demonstrated an effective way for deterministic coupling of quantum emitters with plasmonics. A thick layer of $SiO_2$ was deposited on a semi-insulated Si substrate onto which gold nanopillars are fabricated. Some part of the substrate is deposited with an oxide layer ($Al_2 O_3$). Monolayer $WSe_2$ was fabricated through mechanical exfoliation using adhesive tape and was then transferred onto the substrate containing pillars via dry transfer. Wherever the pillars were present the monolayer took a cone-like structure, inducing a localized strain (Fig.~5). A comparative study shows that the photoluminescence is not dependent on the oxide layer. As interpreted from the spectral intensity map the bright emission spots coincided with the positioning of the nanopillars. These spots showed an enhanced spectral luminescence attributed to the strong resonant coupling to the plasmonic. The single photon property is verified by measuring the autocorrelation function (as low as 0.17). The emissions show strong linear polarization. This polarization if far more pronounced in the extended axis \cite{Luo2017}, as compared to the long axis which is due to coupling between plasmonic in metal and emitters. The way in which monolayer bends around the pillars has a strong impact on polarization \cite{Kern2015}. For square shaped pillars strong field enhancement is observed at the vertical edges of a square that is perpendicular to the electric field polarization, whereas for rod-shaped pillars enhancement is observed in the x-direction and suppression is observed in the y-direction (incident light is polarized in the x-direction). This a very prominent work as it provides scalable cavity electrodynamics with engineered emitters in two-dimensional materials. Similar work has been presented in \cite{Luo2018, Cai2018}. 
 
 Subhojit Dutta et al. \cite{Cai2017} demonstrated the efficient coupling of surface plasmon of the silver nanowire to the localized defects emission site in $WSe_2$ monolayer. The single photon emission is caused due to the strain gradient created in the monolayer when it is placed on the nanowire (Fig.~7(c)). One of the main problems is the precise positioning of the emitter on a nanometer scale, as the surface polaritons decay within a few nanometers \cite{Barnes2003} from the surface. Most of the work reported so far has random positioning \cite{Okamoto2006, Akimov2007, Hartsfield2015, Hoang2015} of emitters with respect to the nanowire. Successful deterministic fabrication \cite{Huck2011, Matsuzaki2017} of emitters sites with respect to plasmons has been done by them, although the method is complex and difficult to scale up. 
 
 In this work, the emitters show no major radiative field enhancement as the electric field vector is perpendicular to the metal surface and directed radially outward from the nanowire. This problem can be solved by using a MIM waveguide, as for two-dimensional materials the dipole moment is in plane \cite{He2015, Chakraborty2015, Koperski2015, Srivastava2015}, tangential to the metal surface. Due to these in-plane components, coupling efficiency is increased effectively. As realized by Subhojit Dutta et al. \cite{Dutta2018} a coupled system of propagating surface plasmons in metal insulator metal waveguide (silver- air- silver) to single photon emitters in $WSe_2$ (Fig.~8(c)). The single photon emission close to the plasmonic modes from the monolayer is due to the strain gradient. For the monolayer, if the in-plane dipole moment is perpendicular to the metal surface then efficient coupling occurs leading to life enhancement. If the monolayer dipole moment is in a plane it can effectively couple to the electric fields in the waveguide (Fig.~8(a-b)). Whereas for nanowires \cite{Cai2017} the dipole moment is radial to the surface, these orthogonal modes do not enable effective coupling. The deterministically active plasmonic circuit can be easily fabricated onto chip \cite{Ajayi2017}. 
 
 \begin{table*}
 	\begin{threeparttable}
 		\caption { \label{tab2} \emph{
 				\textbf{Summary of single photon sources coupled to waveguide.}} }
 		\tabcolsep=0.10 cm
 		
 		\begin{tabular}{lllllllllcccccccccccccccccccccccccccccccccccccc}
 			& \cr
 			\hline
 			Reference & $\beta$ factor  & Purcell   & Propagation  & $g^{(2)}(0)$  & Geometry & FOM \tnote{4} & \cr
 			
 			number &  & factor & length ($\mu$m) & &  &   & \cr                
 			
 			\hline
 			\cite{Faraon2007} & 40, 90 \%    &  50 \tnote{3}  & -  & - & PhC - WC \tnote{4} & - &\cr
 			
 			\cite{Andersen2018} &  85 \% &  18 & $>$120  & - & NVD \tnote{4} - Pl bullseye \tnote{4} & - &\cr
 			
 			\cite{Yang2017} & 0.81 $\pm$ 0.03 & $\approx$ 3.1 $\pm$ 0.3  & -  & - & QD - PhC \tnote{4}  & - &\cr
 			
 			\cite{Chen2015} &  75 \%  &  - & 300   & - & LR-DLSPPWs \tnote{4} - PW & - &\cr
 			
 			\cite{Siampour2017} & 0.58 $\pm$ 0.03 & 5 $\pm$ 1 & 20 $\pm$ 5  & 0.49 $\pm$ 0.02 & NVD \tnote{4}- DLSPPWs \tnote{4}& 83 $\pm$ 15 &\cr
 			
 			\cite{Siampour2018} &   56 $\pm$ 0.03   &  6 $\pm$ 1  & 33 $\pm$ 3  & $<$ 0.5 & GeVD \tnote{4} - DLSPPWs \tnote{4} & 180 &\cr

 			\cite{Bermudez-Urena2015} & 0.42 $\pm$ 0.03 & 2.3 $\pm$ 0.7 & 4.65 $\pm$ 0.48  & $<$ 0.5 & NVD \tnote{4} - VG \tnote{4} & 6.6 $\pm$ 1.5 &\cr
 			
 			\cite{Andryieuski2014} &  26, 15   &  - & 2.5    & - & NA - SWG \tnote{4} & DA \tnote{4} $>$ OCWG \tnote{4}   &\cr
 			
 			\cite{Nielsen2015} & 15 $\pm $ 2 \%  &  - & $\approx$ 10  & - & GSPM  \tnote{4}  & - &\cr
 			
 			\cite{Radko2011} & 9.5 $\pm$ 1.6 \%  \tnote{1}  &  - & 8.0   & - & CPP \tnote{4} - VG \tnote{4} & - &\cr
 			
 			\cite{Bermudez-Urena2017} & 5.5 $\times$ $10^{-4}$ \tnote{2}
 			& - & 19.96  & - & NW \tnote{4} - VG \tnote{4} & - &\cr
 			
 			\cite{Blauth2018} &  8 $\times 10^{-5} $ - 6 $\times$ $ 10^{-3}$ &  15 $\pm$ 3  &   & 0.42 & $WSe_2$ - SWG \tnote{4} & - &\cr
 			
 			\hline
 		\end{tabular}
 		
 		\begin{tablenotes}       
 			\item[1] for 400 nm long taper
 			\item[2] for $R = 0.44$
 			
 			\item[3] when $\beta$ = 90 \%
 			
 			\item[4] Abbreviations: NV-Nitrogen Vacancy in Diamond, VG- V groove, NW- Nanowire, DLSPPWs- Dielectric-loaded surface plasmon polariton waveguide, GeV - Germanium vacancy in diamond, Pl bullseye - Plasmonic bullseye antenna, DA- Dipole antenna, LR-DLSPPWs- Long-range dielectric-loaded surface plasmon-polariton waveguides, PW- Plasmonic Waveguide, WC- waveguide coupler, GSPM- gap surface plasmon, CPP- Channel Plasmon Polaritons, PhC- photonic crystal cavity, OCWG- Open Circuit Wave Guide, NA- Nano Antenna, SWG- Slot Wave Guide. 
 			
 		\end{tablenotes}
 	\end{threeparttable}
 \end{table*}
 \subsection{Conclusion}    
 As indicated in \ref{tab1}, studies have been done to integrated both NON-TMDC and TMDC sources to plasmonics as well as dielectrics. The coupling of NON TMDC sources especially quantum emitters created as a result of vacancies in diamond (NVD - Nitrogen-Vacancy in Diamond, GeV - Germanium Vacancy in Diamond), nanowires (NW), Long-range dielectric-loaded surface plasmon-polariton waveguides (LR-DLSPPWs), gap surface plasmon (GSPM) in metal strip gap film, Channel Plasmon Polaritons (CPP),  Quantum dots (QD created by CdSe/CdS core-shell nanoplatelets coupled to photonic-crystal nanobeam cavity \cite{Yang2017}) has been extensively studied, whereas being comparatively new to the filed the studies showing efficient coupling of TMDC is still limited and have a strong potential for further optimization by strain engineering. Blauth et al. \cite{Blauth2018} coupled $ WSe_2$ to slot waveguide with coupling efficiency as $8 \times 10^{-5} $ to $6 \times 10^{-3} $, Purcell factor of 15 $\pm$ 3 and propagation length increased with increasing physical dimension of the structure. Their efficient coupling still remains an open area of research. The important properties that have to be considered to evaluate an effective coupling have been summarized in \ref{tab2}. Among all the properties coupling efficiency ($\beta$ factor) of quantum emitters to waveguides, decay rate enhancement (Purcell factor), the length to which waveguide modes can travel (Propagation length) are most important. The product of coupling efficiency and Purcell factor normalized by operation wavelength gives FOM (Figure of merit). FOM quantifies the ability of coupling to achieve efficient long-range energy transfer. The second order correlation function at zero time delay ($g^{(2)}(0)$) increases upon coupling with the waveguide as compared to an isolated emitter but is still less than 1 (as expected for a single photon source). A maximum of 90\% efficient coupling has been shown by \cite{Faraon2007} for tilted configuration with two hole separation which decreases to 40\% when the configuration is straight. Nitrogen-vacancy in diamonds coupled to bulls antenna also shows high coupling efficiency of 85\% when the numerical aperture of the objective is 0.9. This configuration also has a very large propagation length $>$ 120 $\mu$m, which further adds to its merits. Until now the longest propagation length has been observed for the Long-range dielectric-loaded surface plasmon-polariton waveguides to plasmonic waveguides that is 300 $\mu$m.In \cite{Andryieuski2014} couples nanoantenna (NA) to slot waveguide (SW) and showed that the $\beta$ factor can be 26 or 15 for two serially connected dipole, and modified bow-tie antenna respectively. Also, the propagation length can increase up to 7 if the leakage losses are eliminated. The DLSPP waveguides have been employed for making directional coupler and can be also used for other components such as cavities to provide a platform for on-chip processing of quantum information. \cite{Siampour2018} demonstrated the way for the integration of an excitation laser, quantum emitter and plasmonic circuit on the same chip. Detection of single plasmons and two-plasmon interference have already been demonstrated on a chip\cite{Heeres2013, Fakonas2014}. With the combination of all these technologies, it will be possible, in the near future, to have all the elements of a quantum plasmonic circuit integrated on a chip. 
 
 \section{Acknowledgment}
 We acknowledge Birla Institute of Technolgy, Mesra, Ranchi for providing research facilities and MHRD, government of India for support through TEQIP - III. We are thankful to Pawan Kumar Dubey and Shubham Adak for their assistance.

\end{document}